%% file: main.tex
\documentclass[a4paper]{jpconf}
\usepackage{graphicx}
\usepackage{xspace}
\usepackage{overpic}
\usepackage{ifthen} 

\newboolean{uprightparticles}
\setboolean{uprightparticles}{false} 

\input{lhcb-symbols-def}
\def\Sigmaplus{\ensuremath{\Sigma^+}\xspace}
 \def\Pp      {\ensuremath{\mathrm{p}}\xspace} 
 \def\PSigma      {\ensuremath{\Sigma}\xspace}

\def\Sigmaplusbar{\ensuremath{\bar \Sigma^-}\xspace}

\def\mpmumu{\ensuremath{m_{p \mu^+ \mu^-}}\xspace}
\def\sigmapxmumu{\ensuremath{\Sigma^+ \to p X^0 (\to \mu^+ \mu^-)}\xspace}
\def\sigmapmumulfv{\ensuremath{\Sigma^+ \to \bar p \mu^+ \mu^+}\xspace}
\def\sigmappiz{\ensuremath{\Sigma^+ \to p \pi^0}\xspace}
\def\sigmappizero{\sigmappiz}
\def\mmumu{\ensuremath{m_{\mu^+ \mu^-}}\xspace}
\def\pizero{\ensuremath{\pi^0}\xspace}

\def\sigmappiz{\ensuremath{\Sigma^+ \to p \pi^0}\xspace}
\def\sigmappizero{\ensuremath{\Sigma^+ \to p \pi^0}\xspace}
\def\kpipipi{\ensuremath{K^+ \to \pi^+ \pi^- \pi^+}\xspace}

\def\NSigmappizero{\ensuremath{(1171 \pm 9)\times 10^{3}}\xspace}

\def\alphafull{\ensuremath{(2.2 \pm 1.2)\times 10^{-9}}\xspace}
\def\expevents{\ensuremath{23 \pm 20}\xspace}
\def\signdefault{\ensuremath{4.1}\xspace}
\def\nsigmadefault{\ensuremath{ 10.2\,^{+\,3.9}_{-\,3.5}}\xspace}

\def\upperlimithcpninety{\ensuremath{1.4 \times 10^{-8}}\xspace}
\def\upperlimithcpninetyfive{\ensuremath{1.7 \times 10^{-8}}\xspace}

\def\brmeasuredsyst{\ensuremath{(2.2\,^{+\,0.9}_{-\,0.8}\,^{+\,1.5}_{-\,1.1})\times 10^{-8}}\xspace}

\def\Sigmaplus{\ensuremath{\PSigma^+}\xspace}

\def\Sigmaplusbar{\ensuremath{\bar \PSigma^-}\xspace}
\def\sigmapmumu{\ensuremath{\PSigma^+ \to \Pp \mu^+ \mu^-}\xspace}
\def\pmumu{\ensuremath{\Pp \mu^+ \mu^-}\xspace}
\def\mpmumu{\ensuremath{m_{\Pp \mu^+ \mu^-}}\xspace}

\def\sigmapxmumu{\ensuremath{\PSigma^+ \to \Pp X^0 (\to \mu^+ \mu^-)}\xspace}
\def\sigmapmumulfv{\ensuremath{\PSigma^+ \to \antiproton \mu^+ \mu^+}\xspace}

\def\sigmappiz{\ensuremath{\PSigma^+ \to \Pp \pi^0}\xspace}
\def\sigmappizero{\sigmappiz}

\def\pizero{\ensuremath{\pi^0}\xspace}

\def\kpipipi{\ensuremath{K^+ \to \pi^+ \pi^- \pi^+}\xspace}

\def\mmumu{\ensuremath{m_{\mu^+\mu^-}}\xspace}
\def\lambdappi{\ensuremath{\PLambda \to \Pp \pi^-}\xspace}
\def\B{\ensuremath{\mathcal{B}}\xspace}

\usepackage{lineno}

\begin{document}
\title{Testing the HyperCP anomaly at LHCb:\\ evidence for the \sigmapmumu decay }

\author{Francesco Dettori \footnote{On behalf of the LHCb collaboration }}

\address{Universit\`{a} degli Studi di Cagliari and INFN, Cagliari, Italy}

\ead{francesco.dettori@cern.ch}

\begin{abstract}
A search for  \sigmapmumu decays with LHCb Run 1 data is presented, 
yielding an evidence for this decay and a measurement of a branching fraction compatible with the Standard Model. 
The HyperCP anomaly in the dimuon invariant mass distribution is not confirmed, 
and its branching fraction central value is excluded at 95\% CL. 
Prospects for the future of this measurement in LHCb are also presented. 
\end{abstract}


The \sigmapmumu decay is a flavour changing neutral current, highly suppressed in the Standard Model. 
It gained attention when the HyperCP collaboration published 
an evidence for this decay~\cite{Park:2005eka}, with a hint of a structure in the dimuon 
invariant mass distribution. 
This decay has been searched recently within the LHCb experiment~\cite{LHCb-PAPER-2017-049}. 
This contribution will discuss this analysis with attention to clarify some points
discussed at this conference while referring the reader to Ref.~\cite{LHCb-PAPER-2017-049} for the rest of the detector and analysis details.

Within the Standard Model, the \sigmapmumu decay~\footnote{The inclusion of the charge-conjugate modes is implied throughout this contribution.}
 is dominated by the long-distance contributions for a predicted
branching fraction of $1.2 \times 10^{-8} < \mathcal{B}(\sigmapmumu) < 10.2 \times 10^{-8}$~\cite{He:2005yn, He:2018yzu}, 
while the short distance contribution is well suppressed at the level of $10^{-13}$. 
Given this large uncertainty, the primary physics interest in this decay comes from the mentioned experimental 
result from the HyperCP experiment.  
The branching fraction measured was $\mathcal{B}(\sigmapmumu) = (8.6^{+6.6}_{-5.4} \pm 5.5) \cdot 10^{-8}$, 
well compatible with the SM predictions. 
However the HyperCP experiment observed that the three clean signal events 
had, within good accuracy, the same dimuon invariant mass value, rather than a phase-space distribution. 
This would point towards and unknown intermediate particle $X^0$ with mass with $m_{X^0} = 214.3 \pm 0.5$ \mevcc. 
The branching fraction, reinterpreted with such an intermediate state, would read 
$\mathcal{B}(\Sigma^+ \to p X^0 ( \to \mu \mu)) = (3.1^{+2.4}_{-1.9}\pm5.5)\cdot 10^{-8}$. 
Albeit limited in statistical significance, this result attracted significant theoretical attention
trying to explain this result~\cite{Gorbunov:2005nu,Demidov:2006pt,He:2005we,Geng:2005ra,Deshpande:2005mb,Chen:2007uv,Xiangdong:2007vv,Mangano:2007gi,Pospelov:2008zw}. 
A considerable experimental effort was put 
in order to search for this particle in other experiments and decays 
\cite{Love:2008aa,Tung:2008gd,Abazov:2009yi,Aubert:2009cp,Hyun:2010an,Abouzaid:2011mi,Ablikim:2011es,Lees:2014xha, Aaij:2013lla, Aaij:2015tna, LHCb-PAPER-2016-052, LHCb-PAPER-2017-038}.
However no other search was done for the exact \sigmapmumu decay prior to the LHCb one. 

The LHCb search for \sigmapmumu decays \cite{LHCb-PAPER-2017-049}, is based on data samples corresponding to an integrated luminosity of 3 \invfb, 
acquired in $pp$ collisions at 7 and 8 TeV centre-of-mass energies in 2011 and 2012 (Run 1). 
The \sigmapmumu candidates are reconstructed in the LHCb detector from three charged tracks, 
with particle identification information consistent with one proton a two muons of opposite sign. 
A standard reconstruction is employed, in particular only \emph{long} tracks are used, 
\ie tracks with hits also in the VELO detector. 
This limits the reconstruction length to decay vertices within about one meter from the primary interaction (PV). 
In the future, other analyses based on different tracking reconstruction could enlarge this to 
about 2 meters, albeit with worse invariant mass resolution. 
The selection of the \sigmapmumu candidates is based on standard kinematic and geometric variables. 
In addition, the normalisation \sigmappizero decay is selected from a track consistent with proton particle identification, 
and a \pizero reconstructed from two photon clusters in the electromagnetic calorimeter. 

Data used to search for \sigmapmumu candidates does not have any special trigger requirement:
all events saved in the considered data set are used in order to maximise the statistics. 
Instead, in order to ease the trigger efficiency evaluation, events used for the normalisation decay are required to belong to events 
triggered by particles independent of the \sigmappizero decay. 
After the mentioned selection, background to \sigmapmumu candidates is composed of combinatorial candidates and 
candidates where a \lambdappi decay, with a pion misidentified as a muon, is combined with an additional muon of the same event. 
Combinatorial background is suppressed through a multivariate selection based on geometric and kinematic variables and  
on isolation information. This is trained with a simulated signal sample and a \sigmapmumulfv (wrong-sign) sample from data to mimic background.
The \lambdappi background is explicitly vetoed by requiring the proton and muon of opposite sign to have a pair invariant mass 
under the $p\pi$ hypothesis, outside a window of 10 MeV from the known $\Lambda$ mass. 
In addition to this, stringent particle identification criteria are used to suppressed both background sources.

\begin{figure}
\begin{overpic}[width = 0.5\textwidth]{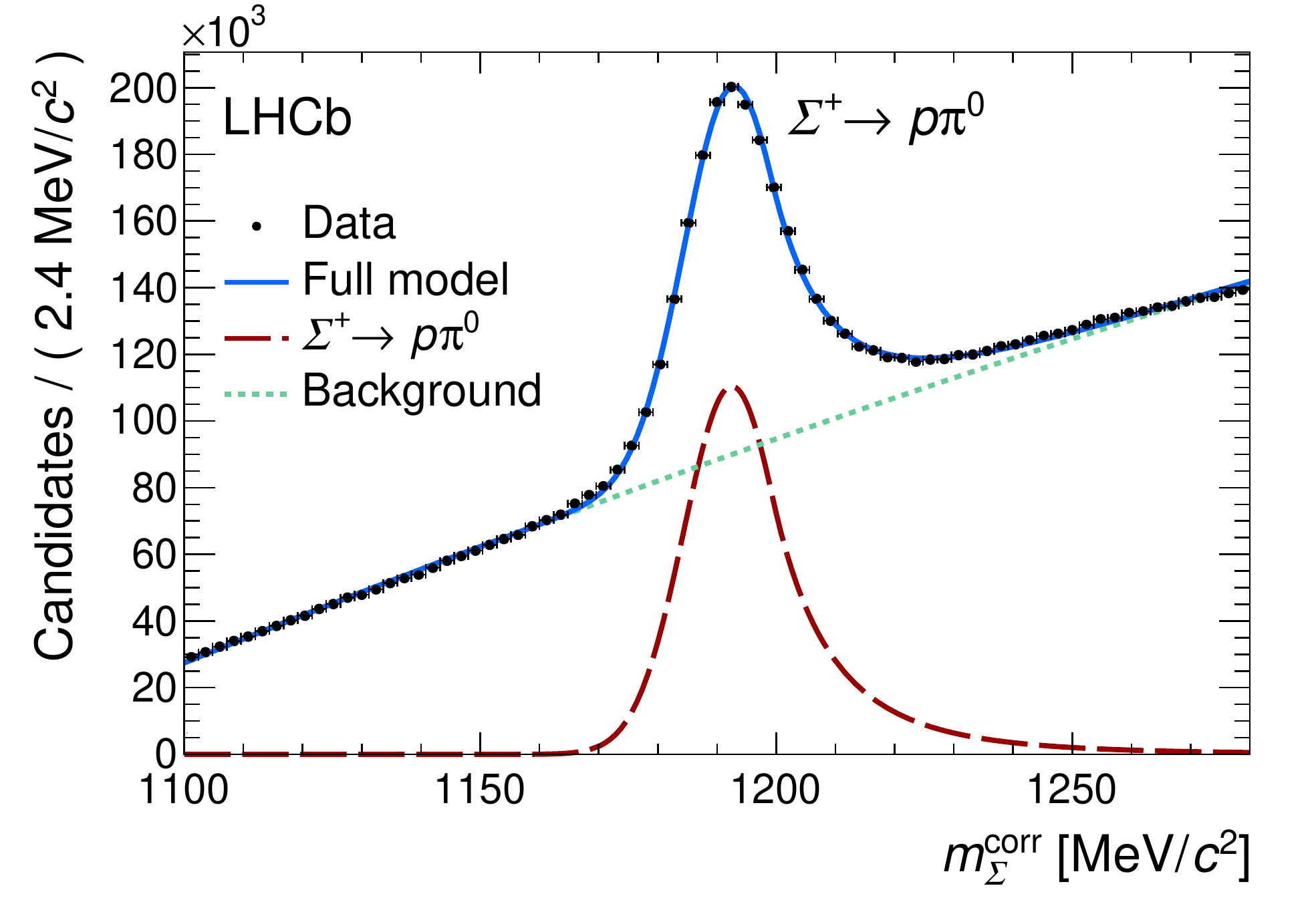}
    \put(80,50){\small (a)}
\end{overpic}
\begin{overpic}[width = 0.5\textwidth]{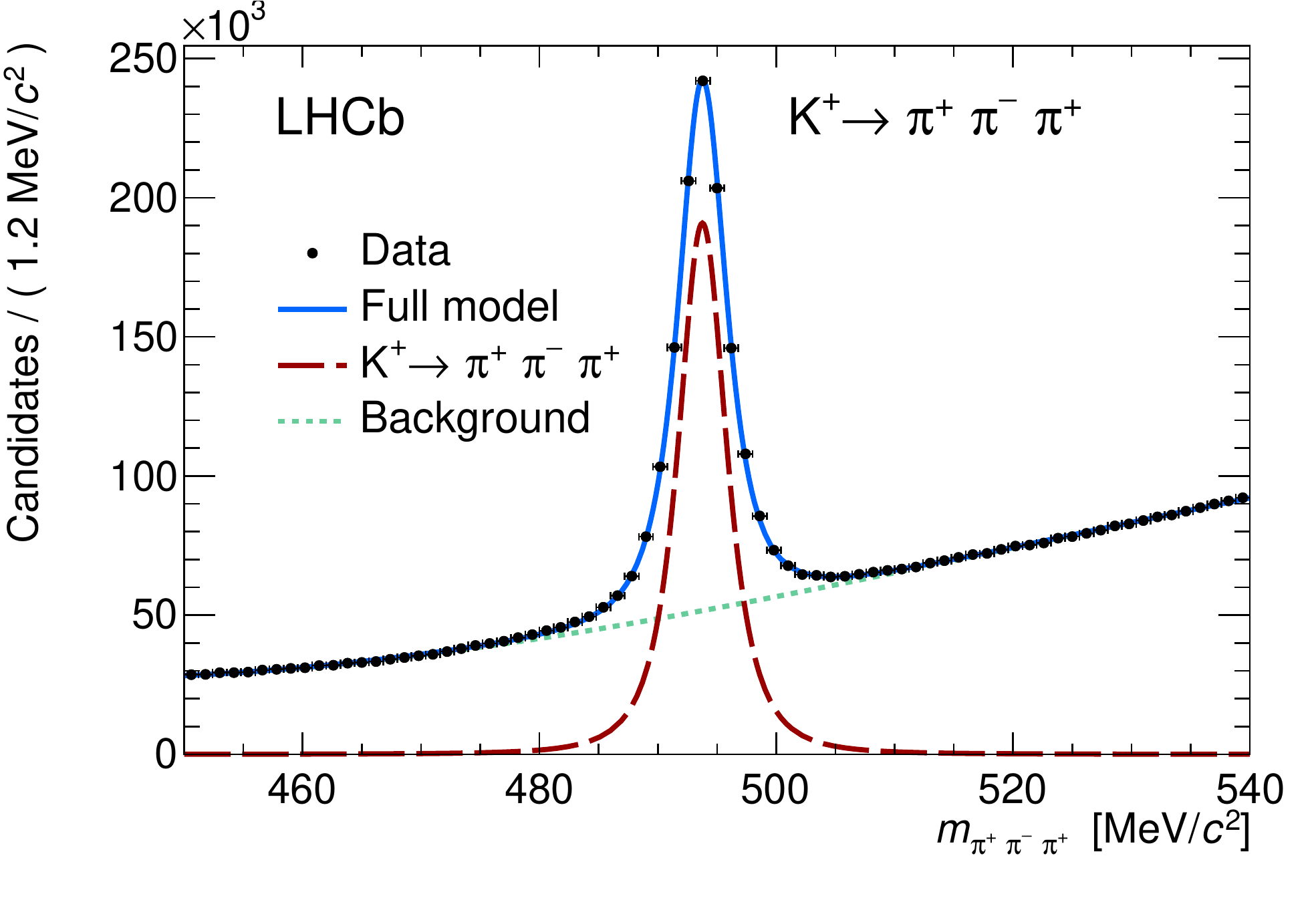}
    \put(80,50){\small (b)}
\end{overpic}
\caption{Invariant mass distribution for (a) \sigmappizero candidates in data, superimposed with the fit and 
(b) \kpipipi candidates in data (see text for details on the fits).}\label{fig:fig1}
\end{figure}

The \sigmapmumu branching fraction is measured by normalising the decay candidates to those of the \sigmappizero decay, 
according to: 
\begin{eqnarray*}
 \B(\sigmapmumu) &=&
\frac{\varepsilon_{\sigmappizero}}{\varepsilon_{\sigmapmumu}}
\frac{ \B(\sigmappizero)}{N_{\sigmappizero}} N_{\sigmapmumu}\\
 &=& \alpha N_{\sigmapmumu}
\end{eqnarray*}
where $\mathcal{B}$, $\varepsilon$ and $N$ are respectively the branching fraction, efficiency and yield of the considered decays. 
The yield of \sigmappizero decays is obtained from a fit to the corrected invariant mass 
$m_{\Sigma}^{\rm{corr}} = m_{p\gamma\gamma} - m_{\gamma\gamma} + m_{\pi^0}$ 
distribution, shown in Fig.~\ref{fig:fig1}. 
The \sigmappizero distribution is described as a Gaussian function with a power
tail on the right side, while the background is described by a modified ARGUS function~\cite{Albrecht:1990am}.
While the reconstruction and selection of \sigmappizero decays is difficult in LHCb, lacking a secondary vertex reconstruction, 
given the very large $\Sigma^+$ production and the large branching fraction ($\sim 50\%$) a good signal can be seen. 
A total of \NSigmappizero \sigmappizero decays are fitted, corresponding to an order of $10^{14}$ $\Sigma^+$ particles in LHCb. 
Selection and reconstruction efficiencies are evaluated in simulations with corrections extracted from data. 
In particular, particle identification efficiencies and trigger efficiencies bring large systematic uncertainties from their calibration on data. 
This is due to samples and techniques being developed in LHCb for higher transverse momentum signals. 
Several of the reconstruction and selection properties are cross-checked with the \kpipipi decay reconstructed 
with a similar selection in data (Fig~\ref{fig:fig1}(b)). 
This will be improved strongly in the future with dedicated samples and tools. 
A single event sensitivity \alphafull is obtained for this analysis, 
meaning that about \expevents \sigmapmumu events are expected in the final sample
considering a typical SM branching fraction of $(5\pm 4 )\times 10^{-8}$.
Considering the HyperCP phase-space branching fraction, this would correspond to about 40 events. 
En passant, we underline, as mentioned earlier, that both $\Sigma^+$ and $\Sigmaplusbar$ are searched for in this analysis 
and that the final sample is composed of about 48\% $\Sigmaplusbar$ anti-baryon candidates.

\begin{figure}
\begin{overpic}[width = 0.5\textwidth]{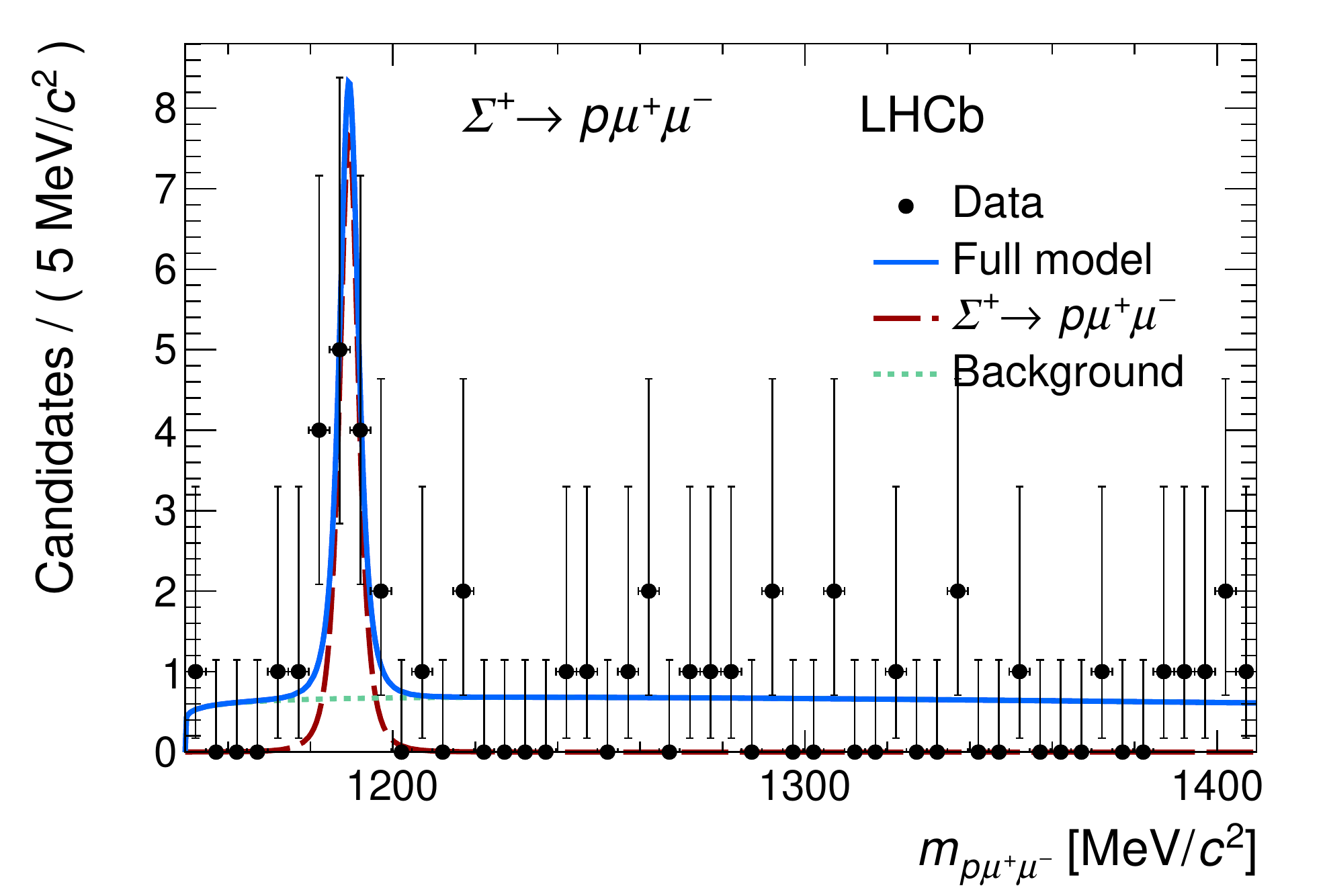}
    \put(85,55){\small (a)}
\end{overpic}
\begin{overpic}[width = 0.5\textwidth]{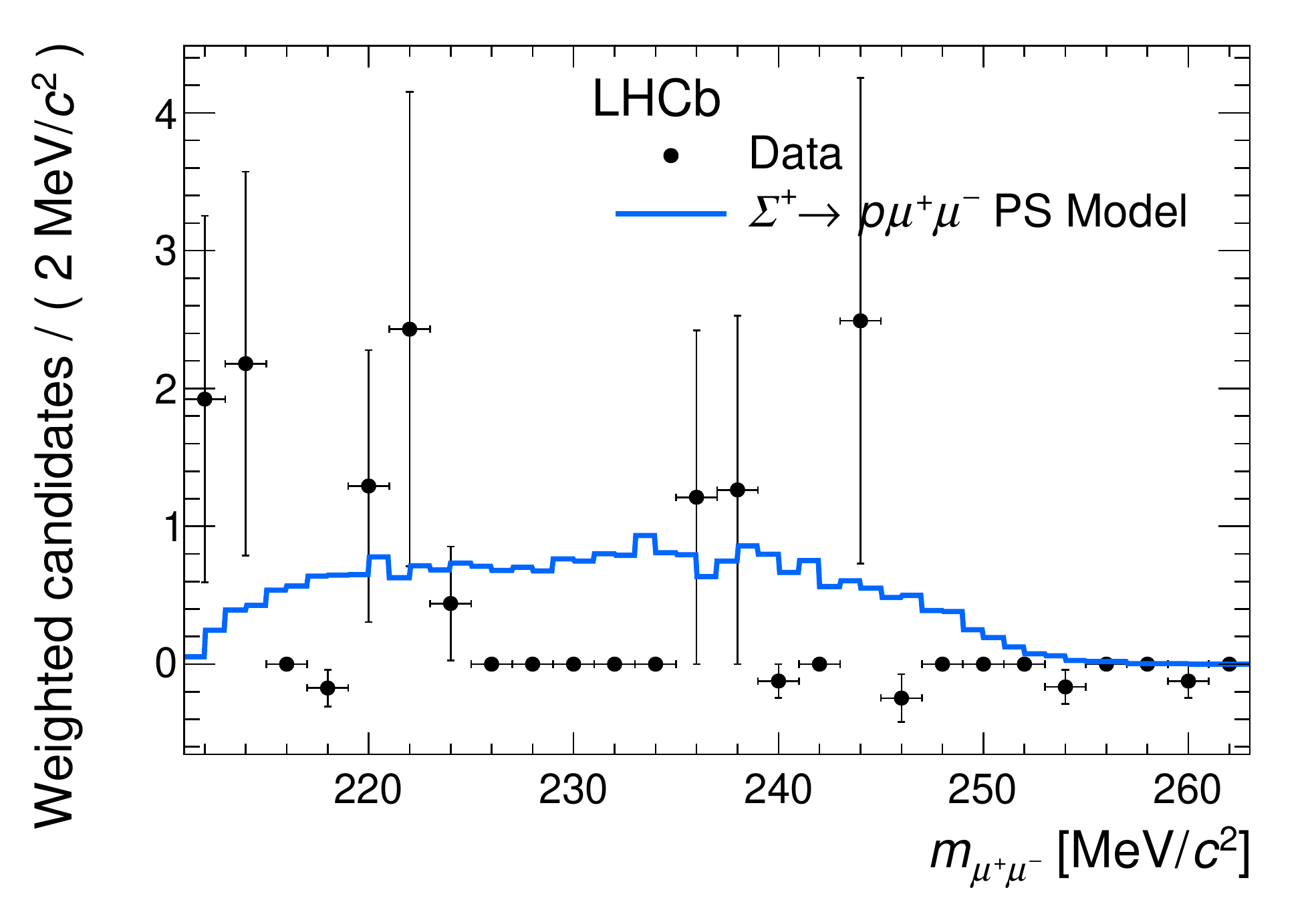}
    \put(85,55){\small (b)}
\end{overpic}
\caption{(a) Invariant mass distribution for \sigmapmumu candidates in data, superimposed with the fit (see text). 
(b) Background subtracted dimuon invariant mass distribution for candidates in the \sigmapmumu final sample.}\label{fig:fig2}
\end{figure}

The final search for \sigmapmumu decays is performed looking for an excess in the \mpmumu invariant mass distribution. 
After the full selection the \lambdappi background is negligible and the only remaining background is of combinatorial nature. 
The invariant mass distribution, shown in Fig.~\ref{fig:fig2}(a),
is described by a modified Argus function for the background and an Hypatia function~\cite{Santos:2013gra}
for the signal. 
An excess of events with respect to the background-only hypothesis is found with a significance of $\signdefault\,\sigma$. 

The fitted signal branching fraction is $\mathcal{B}(\sigmapmumu) = \brmeasuredsyst$, where the first uncertainty is statistical and the second systematic. 
This corresponds to  \nsigmadefault \sigmapmumu decays. The measured branching fraction is compatible with the lower range of the SM prediction. 

The background subtracted dimuon invariant mass distribution for events in the final sample is shown in Figure~\ref{fig:fig2}(b).
This is compared to the distribution from a \sigmapmumu phase space decay. 
It has to be noted that a phase-space distribution or a full amplitude calculation distribution~\cite{He:2018yzu} are very similar.
Incidentally, an HyperCP-like signal would show up as a single peak at 214 \mevcc, with a width of about 0.5 \mevcc coming from the detector resolution. 

In order to search explicitly for an HyperCP-like signal, the fit to the \pmumu invariant mass is repeated restricting to 
events within 1.5 times the resolution from the putative particle ($\mmumu \in [214.3\pm 0.75 ]\mevcc$).
Since no significant signal is found an upper limit on the branching fraction of the resonant decay is set with the $\rm{CL}_{\rm{S}}$ method~\cite{Read:2002hq} 
at ${\mathcal{B}(\sigmapxmumu)<\upperlimithcpninety ~(\upperlimithcpninetyfive)}$ at 90\% (95\%) confidence level. 
This excludes the central value of the HyperCP result in the $X^0$ hypothesis reported above, although not the full range yet given the large uncertainties.

\begin{figure}
\begin{overpic}[width = 0.5\textwidth ]{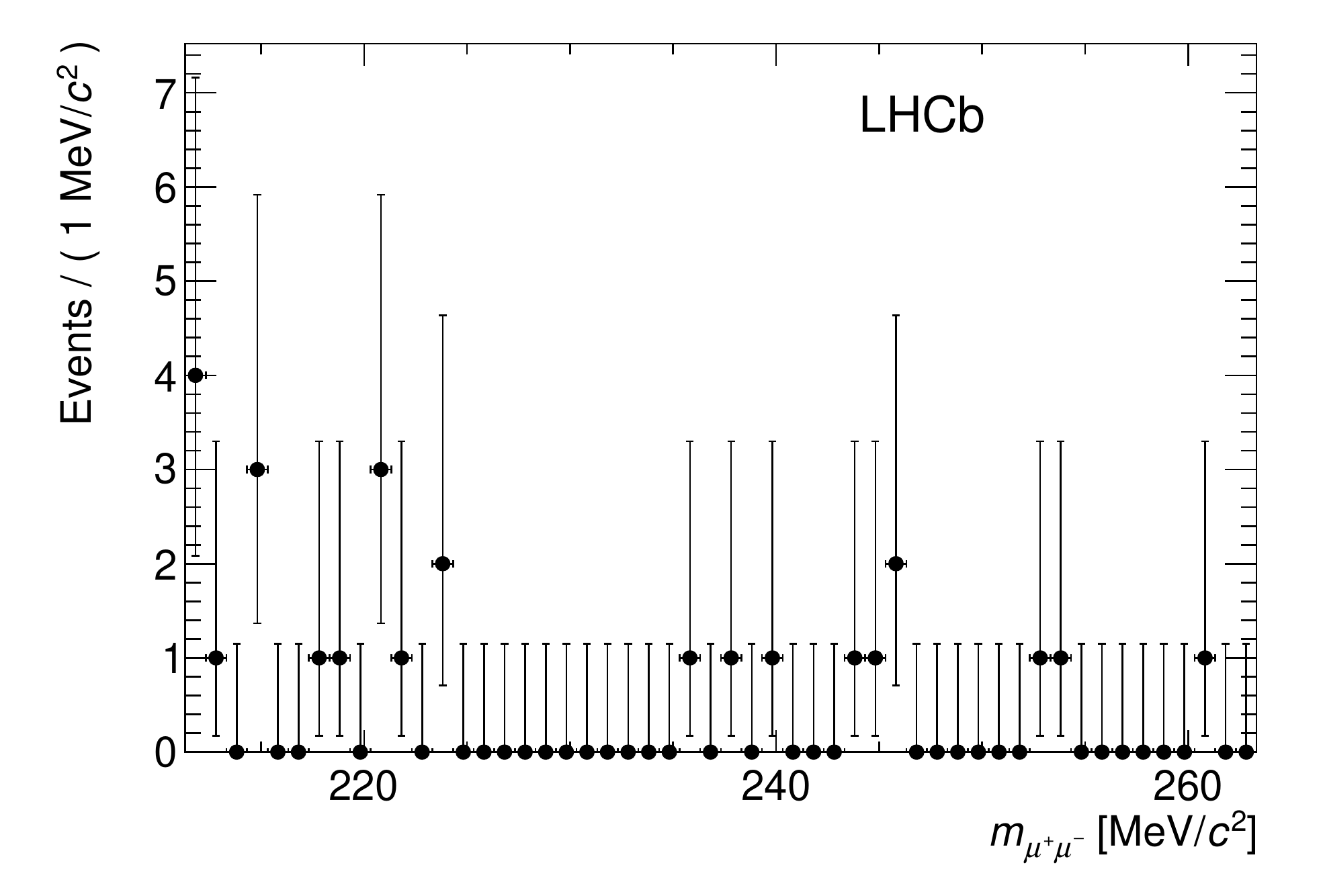} 
    \put(80,45){\small (a)}
\end{overpic}
\begin{overpic}[width = 0.5\textwidth ]{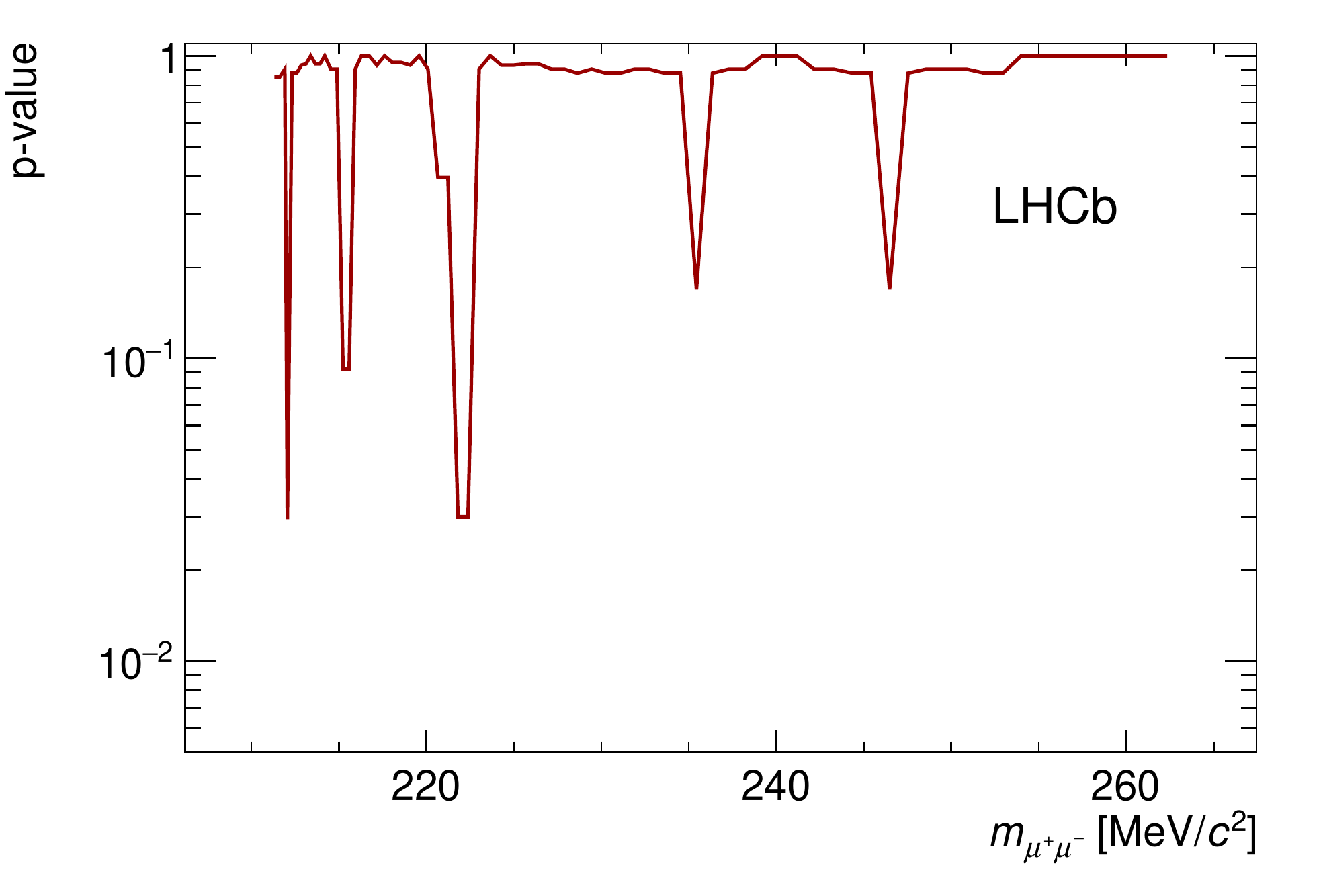} 
    \put(80,45){\small (b)}
\end{overpic}
\caption{(a) Dimuon invariant mass distribution for \sigmapmumu candidates in data (not background subtracted, see text). 
(b) Local p-value in the search for a structure in the dimuon invariant mass distribution.}\label{fig:fig3}
\end{figure}

A more general search for structures in the dimuon invariant mass structure is also done by 
exploiting all the candidates within two times the resolution in the \pmumu invariant mass around the known \Sigmaplus mass. 
The distribution of these candidates (not background subtracted) is shown in Fig.~\ref{fig:fig3}(a). 
In this case it is possible to apply the method developed in Ref.~\cite{Williams:2015xfa} for a generic search, 
to which we refer the reader for details. 
A scan is therefore performed and for each step the putative signal is estimated in a window of $\pm 1.5\times\sigma(m_{\mu^+\mu^-})$
around the considered particle mass,  while the background is estimated from the lower and upper sidebands.
The obtained local p-value of the background-only hypothesis as a function of the dimuon mass is shown
 in Fig.~\ref{fig:fig3}(b) and no significant signal is found. 

The results presented here only analyse Run 1 data of the LHCb experiment. 
During this period no dedicated trigger in any of the LHCb trigger levels was present for this kind of signals. 
Therefore this search is based on very low trigger efficiencies, and is possible owing to the large hyperon production at LHC. 
However for the Run 2 data-taking (2015-2018) different inclusive and exclusive lines have been implemented 
at the first and second software level of the trigger~\cite{Dettori:2297352}.
This will guarantee about an order of magnitude increase in trigger efficiency; this together with another factor ten 
of increase in statistics from luminosity and cross-section in Run 2 allows to foresee a hundred-fold larger dataset 
for the \sigmapmumu decay already collected and soon to be analysed. 
This dataset will ensure not only a high significance observation of this decay, given the current branching fraction, 
but also a precision measurement of this branching fraction. 
This will also be possible thanks to the lower trigger systematic uncertainty that will accompany the dedicated triggers, 
with respect to the current inclusive strategy. 
In addition, it will be possible to study the branching fraction separately for \Sigmaplus and \Sigmaplusbar, allowing 
a search for CP violation in these decays. 
Finally, further observables such as the forward backward asymmetry~\cite{He:2018yzu} will start to be in reach. 
In view of the Upgraded LHCb detector, this will start to become a precision field, not only for this particular decay but also 
for other hyperons at LHCb~\cite{DettoriKaon,RodriguezKaon}.

In conclusion, the search for \sigmapmumu decays at LHCb sees an evidence for this decay in a SM configuration, 
not confirming the HyperCP anomaly, and measures its branching fraction. 
Prospects for the future highlight this analysis just as a starting step of hyperon physics at LHCb.

\section*{References}
\bibliographystyle{unsrt}
\bibliography{main}

\end{document}

%% file: lhcb-symbols-def.tex
\usepackage{xspace} 
\usepackage{upgreek}

\def\MagUp {\mbox{\em Mag\kern -0.05em Up}\xspace}

\ifthenelse{\boolean{uprightparticles}}%
{

 \def\PDelta      {\ensuremath{\Delta}\xspace}                 
 \def\PXi      {\ensuremath{\Xi}\xspace}                 
 \def\PLambda      {\ensuremath{\Lambda}\xspace}                 
 \def\PSigma      {\ensuremath{\Sigma}\xspace}                 
 \def\POmega      {\ensuremath{\Omega}\xspace}                 
 \def\PUpsilon      {\ensuremath{\Upsilon}\xspace}                 
 
 \def\PB      {\ensuremath{\mathrm{B}}\xspace}                 
                  
 \def\PD      {\ensuremath{\mathrm{D}}\xspace}

 \def\PK      {\ensuremath{\mathrm{K}}\xspace}

 \def\Pi      {\ensuremath{\mathrm{i}}\xspace}

 \def\Pp      {\ensuremath{\mathrm{p}}\xspace}

}
{

 \mathchardef\PDelta="7101
 \mathchardef\PXi="7104
 \mathchardef\PLambda="7103
 \mathchardef\PSigma="7106
 \mathchardef\POmega="710A
 \mathchardef\PUpsilon="7107
                  
 \def\PB      {\ensuremath{B}\xspace}                 
                  
 \def\PD      {\ensuremath{D}\xspace}

 \def\PK      {\ensuremath{K}\xspace}

 \def\Pi      {\ensuremath{i}\xspace}

 \def\Pp      {\ensuremath{p}\xspace}

}

\makeatletter
\ifcase \@ptsize \relax
  \newcommand{\miniscule}{\@setfontsize\miniscule{4}{5}}
\or
  \newcommand{\miniscule}{\@setfontsize\miniscule{5}{6}}
\or
  \newcommand{\miniscule}{\@setfontsize\miniscule{5}{6}}
\fi
\makeatother

\DeclareRobustCommand{\optbar}[1]{\shortstack{{\miniscule (\rule[.5ex]{1.25em}{.18mm})}
  \\ [-.7ex] $#1$}}

  \def\Kbar    {{\kern 0.2em\overline{\kern -0.2em \PK}{}}\xspace}

\def\KorKbar    {\kern 0.18em\optbar{\kern -0.18em K}{}\xspace}

  \def\Dbar    {{\kern 0.2em\overline{\kern -0.2em \PD}{}}\xspace}

\def\DorDbar    {\kern 0.18em\optbar{\kern -0.18em D}{}\xspace}

\def\B       {{\ensuremath{\PB}}\xspace}
\def\Bbar    {{\ensuremath{\kern 0.18em\overline{\kern -0.18em \PB}{}}}\xspace}

\def\BorBbar    {\kern 0.18em\optbar{\kern -0.18em B}{}\xspace}

  \def\Y#1S{\ensuremath{\PUpsilon{(#1S)}}\xspace}

\def\proton      {{\ensuremath{\Pp}}\xspace}
\def\antiproton  {{\ensuremath{\overline \proton}}\xspace}

\def\Lbar        {{\ensuremath{\kern 0.1em\overline{\kern -0.1em\PLambda}}}\xspace}
\def\LorLbar    {\kern 0.18em\optbar{\kern -0.18em \PLambda}{}\xspace}

\def\to                 {\ensuremath{\rightarrow}\xspace}

\def\AT#1     {\ensuremath{A_{\mathrm{T}}^{#1}}\xspace}           

\def\C#1      {\ensuremath{\mathcal{C}_{#1}}\xspace}                       
\def\Cp#1     {\ensuremath{\mathcal{C}_{#1}^{'}}\xspace}                    
\def\Ceff#1   {\ensuremath{\mathcal{C}_{#1}^{\mathrm{(eff)}}}\xspace}        
\def\Cpeff#1  {\ensuremath{\mathcal{C}_{#1}^{'\mathrm{(eff)}}}\xspace}       
\def\Ope#1    {\ensuremath{\mathcal{O}_{#1}}\xspace}                       
\def\Opep#1   {\ensuremath{\mathcal{O}_{#1}^{'}}\xspace}                    

\newcommand{\tev}{\ifthenelse{\boolean{inbibliography}}{\ensuremath{~T\kern -0.05em eV}\xspace}{\ensuremath{\mathrm{\,Te\kern -0.1em V}}}\xspace}
\newcommand{\gev}{\ensuremath{\mathrm{\,Ge\kern -0.1em V}}\xspace}
\newcommand{\mev}{\ensuremath{\mathrm{\,Me\kern -0.1em V}}\xspace}
\newcommand{\kev}{\ensuremath{\mathrm{\,ke\kern -0.1em V}}\xspace}
\newcommand{\ev}{\ensuremath{\mathrm{\,e\kern -0.1em V}}\xspace}
\newcommand{\gevc}{\ensuremath{{\mathrm{\,Ge\kern -0.1em V\!/}c}}\xspace}
\newcommand{\mevc}{\ensuremath{{\mathrm{\,Me\kern -0.1em V\!/}c}}\xspace}
\newcommand{\gevcc}{\ensuremath{{\mathrm{\,Ge\kern -0.1em V\!/}c^2}}\xspace}
\newcommand{\gevgevcccc}{\ensuremath{{\mathrm{\,Ge\kern -0.1em V^2\!/}c^4}}\xspace}
\newcommand{\mevcc}{\ensuremath{{\mathrm{\,Me\kern -0.1em V\!/}c^2}}\xspace}

\def\invfb   {\ensuremath{\mbox{\,fb}^{-1}}\xspace}

\def\ms   {\ensuremath{{\mathrm{ \,ms}}}\xspace}

\def\gsim{{~\raise.15em\hbox{$>$}\kern-.85em
          \lower.35em\hbox{$\sim$}~}\xspace}
\def\lsim{{~\raise.15em\hbox{$<$}\kern-.85em
          \lower.35em\hbox{$\sim$}~}\xspace}

\def\tell1  {TELL1\xspace}
\def\ukl1   {UKL1\xspace}

\newcommand{\ie}{\mbox{\itshape i.e.}\xspace}